\documentclass[french,12pt]{article}
\usepackage[english]{babel}
\title{\bf Confinement in $SU(N_c)$ Gauge Theory with a Massive Dilaton}
\author{Mohamed Chabab$^{a,b}$ \footnote{Corresponding author: mchabab@ucam.ac.ma}\hspace{0.1cm}
\quad and \quad Latifa Sanhaji$^a$ \\
\\
{\it $^a)$LPHEA, Physics Department, Faculty of Science - Semlalia}\\
{\it P.O. Box 2390  Cadi-Ayyad University, Marrakech, Morocco.}\\
and
 \\
{\it $^b)$Centro de Fisica Téorica, Departamento de Fisica,}\\
{\it  Universidade de Coimbra, 3004-516 Coimbra, Portugal.} }
\date{}
\begin{document}
\maketitle
\begin{abstract}
\quad Following a recently proposed confinement generating
scenario \cite{Di}, we provide a new string inspired model with a
massive dilaton and a general dilaton-gluon coupling. By solving
analytically the equations of motion, we derive a new class of
confining interquark potentials, which includes most of the QCD
motivated potential forms given in the literature.
\end{abstract}

\newpage

\section{Introduction}
To describe the confinement of quarks and gluons, several low
energy effective models have been proposed. The most popular ones
are : Color dielectric models \cite{JTC,FL,BR}, the constituent
models with non relativistic quark and a confining potential
\cite{Gr}, and the dual Landau-Ginzberg Model \cite{Ho}. Recently,
the extension of gauge field theories by inclusion of dilatonic
degrees of freedom has evoked considerable interest.
Particularly, dilatonic Maxwelle and Yang-Mills theories which,
under some assumptions, possess stable and finite energy
solutions \cite{CT}. Indeed, in theories with dilaton fields, the
topoligical structure of the vacuum is drastically changed
compared to the non dilatonic ones. It is therefore of great
interest to investigate the vacuum solutions induced by
r-dependent dilaton field, through a string inspired effective
theory which may reproduce the main features of QCD, in
particular, the quark confinement. Recall that the dilaton is an
hypothetical scalar particle appearing in the spectrum of string
theory and Kaluza-Klein type theories \cite{GSW}. Along with its
pseudo scalar companion, the axion, they are the basis of the
discovery F-theory compactification \cite{Va} and of the
derivation of type IIB self duality \cite{Se}. The main features
of a dilaton field is its coupling to the gauge fields through
the Maxwell and Yang-Mills kinetic term. In particular, in string
theory, the dilaton field determines the strength of the gauge
coupling at tree level of the effective action. In this context,
Dick \cite{Di} observed that a superstring inspired coupling of a
massive dilaton to the 4d $SU(N_c)$ gauge fields provides a
phenomenologically interesting interquark potential $V(r)$ with
both the Coulomb and confining phases. The derivation performed
in \cite{Di} is phenemenologically attractive since it provides a
new confinement generating mechanism. In this context, a general
formula of a quark-antiquark potential, which is directly related
to the dilaton-gluon coupling function, has been obtained in
\cite{Ch1}. The importance of this formula is manifest since it
generalizes the Coulomb and Dick potentials, and it may be
confronted to known descriptions of the confinement,
particularly, those describing the complex structure of the
vacuum in terms of quarks and gluons condensates. Moreover a
generalized version of Dick model with both a massive and
massless dilaton has been proposed in \cite{SW}\footnote{there is
a missing factor $(q)^\frac{1}{1+4\delta}$ in second term of
Eq.(12) and Eq.(13). Also, the second term of Eq.(14) should be
multiplied by $q^\frac{3}{4}$.}.

In this paper, we shall propose a new effective coupling of a
massive dilaton to chromoelectric and chromomagnetic fields
subject to the requirement that the Coulomb problem still admits
an analytic solution. Our main interest concerns the derivation
of a new family of confining interquark potentials. As a by
product, we shall set up a theoretical basis to various QCD
motivated quark potentials used    in the literature. The later
would gain in credibility if they can emerge from low energy
effective theories.

The plan of this work is as follows: In section 2, we will
develop our model and derive the main equations of motion.
Particularly, emphasis will be put on the equations of a massive
dilaton in the asymptotic regime. The latter should show the long
range behaviour of the solutions, and consequently is connected
with the confining phase. Section 3 will be devoted to the
existence of analytical solutions from which we shall extract a
new class of interquark potentials whose magnitude grows with the
separation between the quark and antiquark. The main features of
these potentials will be presented along with their connection to
some popular phenomenological ones. Finally our conclusion will be
drawn in section 4.

\section{The model}
We propose an effective field theory defined by the general
Lagrangian:
\begin{eqnarray}
{\cal L}({\phi},A)&=&
-\frac{1}{4F({\phi})}{G_{{\mu}{\nu}}^a}{G^{{\mu}{\nu}}_a}
+\frac{1}{2}\partial_\mu \Phi \partial^\mu  -V(\phi) +J_a^\mu
A_\mu^a
\end{eqnarray}
where the coupling function $F(\phi)$ depends on the dilaton
field and $V(\phi)$ denotes the non perturbative scalar potential
of $\phi$. $G^{\mu \nu}$ is the field strength in the
language of 4d gauge theory.\\
Several forms of the function $F(\Phi)$ appeared in different
theoretical frameworks: $F(\Phi)=e^{-k\frac{\Phi}{f}}$ as in
string theory and Kaluza-Klein theories \cite{GSW};
$F(\Phi)=\frac{\Phi}{f}$ in the Cornwall-Soni model
parameterizing the glueball-gluon coupling \cite{CS, DF}. As to
Dick model, $F(\Phi)$ is given by $F(\Phi)=k+\frac{f^2}{\Phi^2}$.
The constant f is a characteristic scale of the strength of the
dilaton/glueball-gluon. By using the formal analogy between the
Dick problem and the Eguchi-Hansen one \cite{GIOT}, we noted in
\cite{Ch1} that $f$ is similar to the $4d N=2$ Fayet-Illioupoulos
coupling in the Eguchi-Hansen model. It may be interpreted as the
breaking scale of the $U(1)$ symmetry rotating the dilaton field.

Now, to analyze the problem of the Coulomb gauge theory augmented
with dilatonic degrees of freedom in (1), we proceed as follows:
first, we consider a point like static Coulomb source which is
defined in the rest frame by the current:
\begin{eqnarray}
J_a^\mu =g \delta (r) C_a \nu_0^\mu =\rho_a \nu_0^\mu
\end{eqnarray}
$C_a$ represents the expectation value of the $SU(N)$ generator
$\chi_a$ for a normalized spinor in Coulomb space. They satisfy
the algebraic identity:
\begin{eqnarray}
\sum_{a=1}^{N^2 -1} C_a^2 =\frac{N_c -1}{2N_c}
\end{eqnarray}
The equations of motions, inherited from the model (1) and
emerging from the static configuration (2) are given by:
 \begin{eqnarray}
 \left[ D_\mu , F^{-1} (\Phi ) G^{\mu\nu}\right] = J^\nu
\end{eqnarray}
 and
\begin{eqnarray}
 \partial_\mu \partial^\mu \Phi = -\frac{\partial
 V(\Phi)}{\partial\Phi}-\frac{1}{4} \frac{\partial F^{-1}(\Phi)}{\partial
 \Phi}G_a^{\mu\nu}G_a^{\nu\mu}
 \end{eqnarray}

By setting $G_a^{0i} = E^i \chi_a =-\nabla^i \Phi_a$, we obtain,
after some straightforward algebra, the simplified expressions:
\begin{eqnarray}
\frac{d\Phi_a}{dr} =r^{-2}F(\Phi(r))\left(\frac{-g}{4\pi}
C_a\right)
  \end{eqnarray}
  \begin{eqnarray}
  \Delta \Phi = \frac{\partial V}{\partial\Phi} -\frac{\widetilde{\alpha}}{r^4} \frac{\partial F}{\partial\Phi}
\end{eqnarray}
with $\widetilde{\alpha} =\frac{g^2}{32\pi^2} \left( \frac{N_c -1
}{2N_c}\right)$ We then derive the important formula of
\cite{Ch1, Ch2},
 \begin{eqnarray}
\Phi_a (r) = \frac{-g C_a}{4\pi}\int dr \frac{F(\Phi(r))}{r^2}
\end{eqnarray}
which shows that the quark confinement appears if the following
condition is satisfied:
\begin{eqnarray}
\lim_{r\to \infty} r F^{-1}(\Phi(r)) = finite
\end{eqnarray}
Then, the interquark potential reads as,
\begin{eqnarray}
U(r)&=&\Phi_a (r) \left( \frac{-g}{4\pi} C^a\right) \nonumber\\
&=&2\widetilde{\alpha}_s \int \frac{F(\Phi(r))}{r^2} dr
\end{eqnarray}

At this stage, note that the effective charge is defined by,
$$Q^a_{eff}(r)=\left(g\frac{C_a}{4\pi}\right) F(\Phi(r))$$ thus the
chromo-electronic field  takes the usual standard form: $$E_a
=\frac{Q^a_{eff}(r)}{r^2}$$Therefore, it is the running of the
effective charge that makes the potential stronger than the
Coulomb potential. Indeed if the effective charge did not run, we
recover the Coulomb spectrum.

To solve the equations of motion (6) and (7), we need to fix two
of the four unknown quantities $\Phi(r)$, $F(\Phi)$, $V(\Phi)$
and $\Phi_a(r)$ in our model. We set $V(\Phi)$ to $V(\Phi)
=\frac{1}{2} m^2 \Phi$ and we introduce a new coupling function:
$$F(\Phi)= \left(1-\beta\frac{\Phi^2}{f^2}\right)^{-n}$$

Then the equation (7) becomes:
\begin{eqnarray}
\Delta\Phi =m^2 \Phi -2n\frac{\widetilde{\alpha}_s}{r^4}\left(
\left(1-\frac{\beta\Phi^2}{f^2}\right)^{-(n+1)}\right)
\frac{\beta\Phi}{f^2}
\end{eqnarray}
This equation is very difficult to solve analytically. However
since we are usually interested by the large distance behaviour
of the dilaton field and its impact on the Coulomb problem, an
analytical solution of (11) in the asymptotic regime is very
satisfactory. Indeed, it is easily shown that the following
function:
\begin{eqnarray}
\Phi =\left[\frac{f^2}{\beta}
-\left(\frac{\beta}{f^2}\right)^\frac{-n}{n+1}
\left(\frac{2n\alpha_s}{m^2}\right)^\frac{1}{n+1}
\left(\frac{1}{r}\right)^\frac{4}{n+1}\right]^\frac{1}{2}
\end{eqnarray}
solves (11) at large $r$. Therefore, thanks to the master formula
(11), we derive the potential,
\begin{eqnarray}
\Phi_a (r)=-\frac{g C_a}{4\pi}\left( \frac{2n\alpha_s}{m^2
f^2}\right)^\frac{-4n}{n+1} \frac{n+1}{3n-1} r^{\left(
\frac{3n-1}{n+1} \right)}
\end{eqnarray}
By imposing the condition (9), we obtain a family of confining
interquark potentials if $n\geq\frac{1}{3}$. If moreover, we use
the criterion of Seiler \cite{Se} which states that the magnitude
of confining potentials, can not grow more rapidly than linear,
then the values of $n$ are constrained to the range $n\leq 1$.
Therefore the confinement in our model (1) appears for the
coupling function $\frac{1}{F(\Phi)}$ with $n \in
\left[\frac{1}{3},1\right]$. Such class of confining potentials
is very attractive. Indeed, by selecting specific values of $n$,
we may reproduce several popular QCD motivated interquark
potentials: Indeed if $n=1$, we recover the confining linear term
of Cornell potential \cite{Ei}. Martin's potential $(V(r)\sim
r^{0.1} )\cite{Ma}$ corresponds to $n=\frac{11}{29}$, while
Song-Lin interquark potential \cite{SL} and Motyka-Zalewski
potential \cite{MZ}, with a long range behaviour scaling as
$\sqrt{r}$, are obtained by setting $n$ to $\frac{3}{5}$. Turin
potential \cite{Li} is recovered for $n=\frac{5}{9}$. We see then,
that these phenomenological potentials, which gained credibility
only through their confrontation to the hadron spectrum, can now
have a theoretical basis since they can be derived from the low
energy effective theory.

\section{Conclusion}
In this paper we have found a family of electric solutions
corresponding to a string inspired effective gauge theory with a
massive dilaton varying with $r$ and a new coupling function
$F(\Phi)= \left(1-\beta\frac{\Phi^2}{f^2}\right)^{-n}$. By
constraining the values of n by both the Seiler criterion and by
condition of Eq.(9) we have shown the existence of a class of
confining interquark potentials. The latter are
phenomenologically interesting since they reproduce, through
selecting specific values of $n$, several QCD motivated
potentials which successfully describe meson and baryon spectra.
Clearly these popular potentials would gain in credibility since
they emerge from an low energy effective theory, and at the same
time fit well the hadron spectrum.

\section*{Acknowledgements}
One of the authors (M.C) is deeply grateful to the Centro de
Fisica Teorica for its warm hospitality in Coimbra. He wishes to
thank prof.J. da Providencia for the valuable discussions and
comments.

This work is supported by the convention
CNRST-Morocco/GRICES-Porugal, grant 681.02/ CNR and by the
PROSTARS III program D16/04.

\end{document}